\documentclass{emulateapj}
\slugcomment{Accepted for publication in the ApJ}
\lefthead{Laha et al.}
\usepackage{color}
\usepackage{rotating}
\usepackage{lscape}

\def\unit #1{\,{\rm #1}}

\newcommand\kms{\rm \,\unit{km\,s^{-1}}}

\newcommand\cmsqi{\rm \,\unit{cm^{-2}}}

\newcommand\cmcubei{\rm \,\unit{cm^{-3}}}

\newcommand\kev{\rm \,\unit{keV}}

\newcommand\funit{\rm \,erg\,cm^{-2}\,s^{-1}}

\newcommand\lunit{\rm \,erg \,s^{-1}}

\newcommand\xiunit{\rm \,erg\,cm\,s^{-1}}

\newcommand\msun{M_{\odot}}

\newcommand\nh{\rm N_{H}}

\newcommand\nhwa{N^{\rm WA}_{\rm H}}

\newcommand\ks{\, \rm ks}

\newcommand\dc{\, \Delta\chi^2}

\newcommand\cd{\,\rm \chi^2/dof}

\newcommand\ev{\unit{\, eV}}

\newcommand\chandra{{\it Chandra \,}}

\newcommand\xmm{{\it XMM-Newton \,}}

\usepackage[normalem]{ulem}  

\usepackage{graphicx}

\begin{document}

\title{The effect of UV/Soft X-ray excess emission on the warm
  absorber properties of Active Galactic Nuclei -- A case study of
  IRAS~13349+2438.}

\author{Sibasish Laha\altaffilmark{1}, Gulab
  C. Dewangan\altaffilmark{1}, Susmita Chakravorty\altaffilmark{2,}\altaffilmark{ 3}, Ajit
  K. Kembhavi\altaffilmark{1}}\altaffiltext{1}{{Inter University
    Centre for Astronomy and Astrophysics}; {\tt email:
    laha@iucaa.ernet.in ; gulabd@iucaa.ernet.in} }\altaffiltext{2} {Harvard University, Department of Astronomy}
  \altaffiltext{3} {Harvard Smithsonian centre for Astrophysics, 60 Garden street, Cambridge, MA 02138, USA}

\begin{abstract}

  The UV to X-ray continuum of active galactic nuclei (AGN) is
  important for maintaining the ionisation and thermal balance of the
  warm absorbers (WAs). However, the spectra in the sensitive energy
  range $\sim \,13.6 -300 \ev$ are unobservable due to Galactic
  extinction. Moreover, many AGN show soft X-ray excess emission of
  varying strength in the $0.1-2\kev$ band whose origin is still highly debated. This soft-excess connects to the UV bump in the unobserved region of $13.6 -300 \ev$. Here we investigate the effect of the assumed physical model for the soft-excess on the flux of the unobserved part of the spectrum and its effect on the WA properties. We perform a case study using the \xmm{} observations of the bright Seyfert 1 galaxy IRAS~13349+2438 with WA features. The two different physical models for the soft excess: blurred Compton reflection from an ionised disk, and, optically thick thermal Comptonisation of the disk photons, predict different fluxes in the unobserved energy range. However the current X-ray data quality does not allow us to distinguish between them using derived WA parameters. This, in turn, implies that it is difficult to determine the origin of the soft-excess emission using the warm absorber features.

\end{abstract}

\vspace{0.5cm}


\section{INTRODUCTION}

About half of type I active galactic nuclei (AGN) show soft X-ray
absorption features due to partially ionised material along our line
of sight and intrinsic to the source \citep{1994MNRAS.268..405N,
  1997ASPC..128..173R, 1998ApJS..114...73G, 2005A&A...431..111B,
  2005A&A...432...15P} . Such X-ray absorbing clouds first detected by
\cite{1984ApJ...281...90H} using {\it Einstein} data, have been named
partially ionised absorbers or ``warm absorbers''. The
availability of high resolution grating X-ray spectra with \xmm{} and
\chandra{} in the last decade improved enormously our knowledge of
these discrete warm absorption and warm emission features in AGN
spectra.

The warm absorber (hereafter WA) clouds give rise to narrow absorption
lines and edges in the spectrum, from various ionisation stages of
several elements \citep[see e.g.,][]{2000A&A...354L..83K,
  2000ApJ...535L..17K,2005A&A...431..111B}.  Absorption features from H-like and He-like
ions of C, N, O, Ne and lower ionisation states of Fe (including the
unresolved transition array (UTA) at $\sim 0.7 \kev $) are most
prominent in the soft X-rays. These lines
and edges are sensitive to photons in the energy range of $13.6\ev- 2
\kev$, and the ionisation structure of the WA clouds depends on the
shape and strength of the AGN continuum in that energy range. As such,
these lines serve as important diagnostics of the ionisation structure
and kinematics of the gas, as well as a probe of the continuum shape.

To model these lines and edges in a dataset and determine the
ionisation phase (via an ionization parameter $\xi$) and column
density $\nh$ of the cloud, there are a few photo-ionisation codes in
vogue, for example, CLOUDY \citep{1998PASP..110..761F}, XSTAR \citep{2004ApJS..155..675K}, etc. These codes require as inputs the ionizing
continuum from the source, cloud density and cloud metallicity, among other
parameters, to generate a grid in $\xi$ and $\nh$. The ionizing continuum plays a very important role in determining the parameters of the cloud.

However, it is not always easy to have an accurate estimate of the ionizing
continuum as emitted by the central engine and as seen by the WA. Due
to Galactic {neutral absorption} a portion of the ionizing continuum
($13.6-300\ev$) becomes unobservable to us. This is the region where the two most important parts of an AGN spectral energy distribution (SED), the Big Blue Bump (BBB) and the soft X-ray excess (SE) join. The BBB {(peaking at $\sim 1-30\ev)$} is believed to have its origin in the
accretion disk, as thermal multi-temperature blackbody
emission \citep{1973A&A....24..337S}. The SE, on the other
hand, is excess emission at energies $< 2 \kev$ over a powerlaw
extending to high energies. Till date there has been no consensus on
the physical origin of the SE and it can be well described by
many prevalent models such as single or multiple blackbodies, high
temperature diskblackbody, optically thick thermal Comptonization,
blurred reflection from partially ionized disk etc \citep[see e.g.,][]{2005MNRAS.358..211R,2012MNRAS.420.1848D}. The different models when used to describe the SE can predict different fluxes when extrapolated to the unobserved portion of the SED; the photons from this part of the spectrum are particularly important for warm absorber clouds.  So a physical description of the soft-excess becomes imperative to describe the broad band SED. Moreover, the energy range of the soft-excess ($0.3-2\kev$) is
where the WA features are mainly found. Hence the properties of the WA
are likely to depend on the way the SE is modeled.

\cite{1999ApJ...517..108N} have studied the properties of the transmitted spectra of a gas illuminated by a flat and a steep X-ray spectrum. They found that different X-ray continua produce distinctly different ionisation structure in a cloud resulting in different absorption features in the energy range $0.1-2 \kev$.  \cite{2012A&A...542A..30M} have investigated the effect of the uncertainties in the construction of the SED of the Seyfert 1.8 galaxy ESO~113-G010 on its WA, as it is intrinsically obscured. The uncertainties in the IR and the UV parts of the SED were tested and were found to affect the thermal stability of each phase of the detected warm absorbers. \cite{2013MNRAS.tmp..803L} have studied the warm absorbers of the source IRAS~13349+2438 using \chandra data. They have found that the presence of the UV bump in the SED creates an increased number of stable phases in the Stability-curves and thereby favor a continous distribution of ionisation states in pressure equilibrium.

In this paper we use the different physical models describing the observed X-ray and the UV data and predict the fluxes in the unobserved part of the continuum ($13.6-300\ev$). Further we investigate the effect of these different continua on the WA properties, using a case study of an \xmm{} observation of a bright Seyfert 1 galaxy
IRAS~13349+2438, known to show strong WA features
\citep{2001A&A...365L.168S}. We also investigate how the different shapes of the UV and
soft X-ray continua affect the WA properties of the source. We further carry out an extensive stability curve analysis for the different SEDs.


The paper is organised as follows. Section 2 deals with the different
ionizing continua and their effect on the X-ray spectrum. Section 3 deals with the case study of
IRAS~13349+2438 and describes X-ray observation, modeling of continuum
and WA absorbers. This section also describes the construction of an appropriate ionizing continuum for IRAS~13349+2438 and also investigates the effect of different ionizing continua on the observed
WA properties and the stability curve analysis. Section 4 discusses
the results followed by conclusions in Section 5. Throughout this work we have used the cosmological parameters $\rm H_o= 71 \, \rm km\,s^{-1}\,Mpc^{-1}$,\,$\Omega_m =0.27$,\, $\Omega_{\Lambda}=0.73$ to calculate distance.



\section{Ionizing continuum and the warm absorbers}
\label{sec:sed}

The level of ionisation in a warm absorber cloud can be characterised by the ionisation parameter $\xi = L/nr^2$, where $L$ is the ionizing
luminosity between 1 and 1000 Ryd, $n$ is the hydrogen number density, and
$r$ distance between the ionizing source and the illuminated face of the cloud \citep{1969ApJ...156..943T}. This parameter is defined for hydrogen at the cloud surface facing the ionising radiation. The ionisation structure of the cloud on the other hand, determined by the relative ionic abundances in the photo-ionised gas, depends on the shape of the incident spectrum. Clouds having same ionisation parameter $\xi$ can show different absorption features when illuminated by different SEDs. \cite{1999ApJ...517..108N} have found that the steep sloped NLSy1s lack the presence of strong absorption edges of Oxygen and other elements whereas the flat sloped Seyfert 1 galaxies show strong presence of K$\alpha$ and K$\beta$ resonance absorption lines from H-like and He-like ions of C, O, Ne, etc. The continuum strength in the soft X-rays therefore play a crucial role in determining the nature of ionic structure of the WA cloud. 
This has important implication for the extent of
  ionization in WA clouds, as the photoelectric absorption cross-sections for
  various elements are generally higher at lower energies above their
  K-edges.  Thus, a continuum with strong extreme ultra-violet and soft
  X-ray excess emission will produce more ionization compared to a
  flat continuum with strong hard X-rays. To demonstrate these
  effects, we have created warm absorber models for four different
  ionizing continua with varying UV and soft X-ray excess emission.
The choice of these continua are driven by the case study we
perform on the source IRAS13349+2438 in later sections. The four ionizing continua are as follows.
\begin{enumerate}
  
\item NLSy1 continuum -- This continuum is typical of Seyfert 1
  galaxies and consists of an X-ray power-law (photon index $\Gamma \sim 2$), soft
  X-ray excess below $2\kev$ described as a blackbody ($kT_{BB} \sim
  85\ev$), and a multicolor accretion disk blackbody (BBB) characterised by
  an inner disk boundary temperature $kT_{in}\sim 4\ev$. This is the
  UV-X-ray continuum obtained from \xmm{} observation of
  IRAS~13349+2438 (see Sect.~\ref{subsec:xray-cont}) and the BBB as characterised by \cite{2013MNRAS.tmp..803L}  (see Sect. \ref{subsec:BBB}).

\item NLSy1 continuum without the BBB.
\item NLSy1 continuum without the SE.
\item NLSy1 continuum without the BBB and the SE.

\end{enumerate}
The continua 2, 3 and 4 are generated to investigate the effect of various parts of the
ionising continuum on the warm
absorber clouds.


\begin{figure*}
  \centering
  \includegraphics[width=8cm]{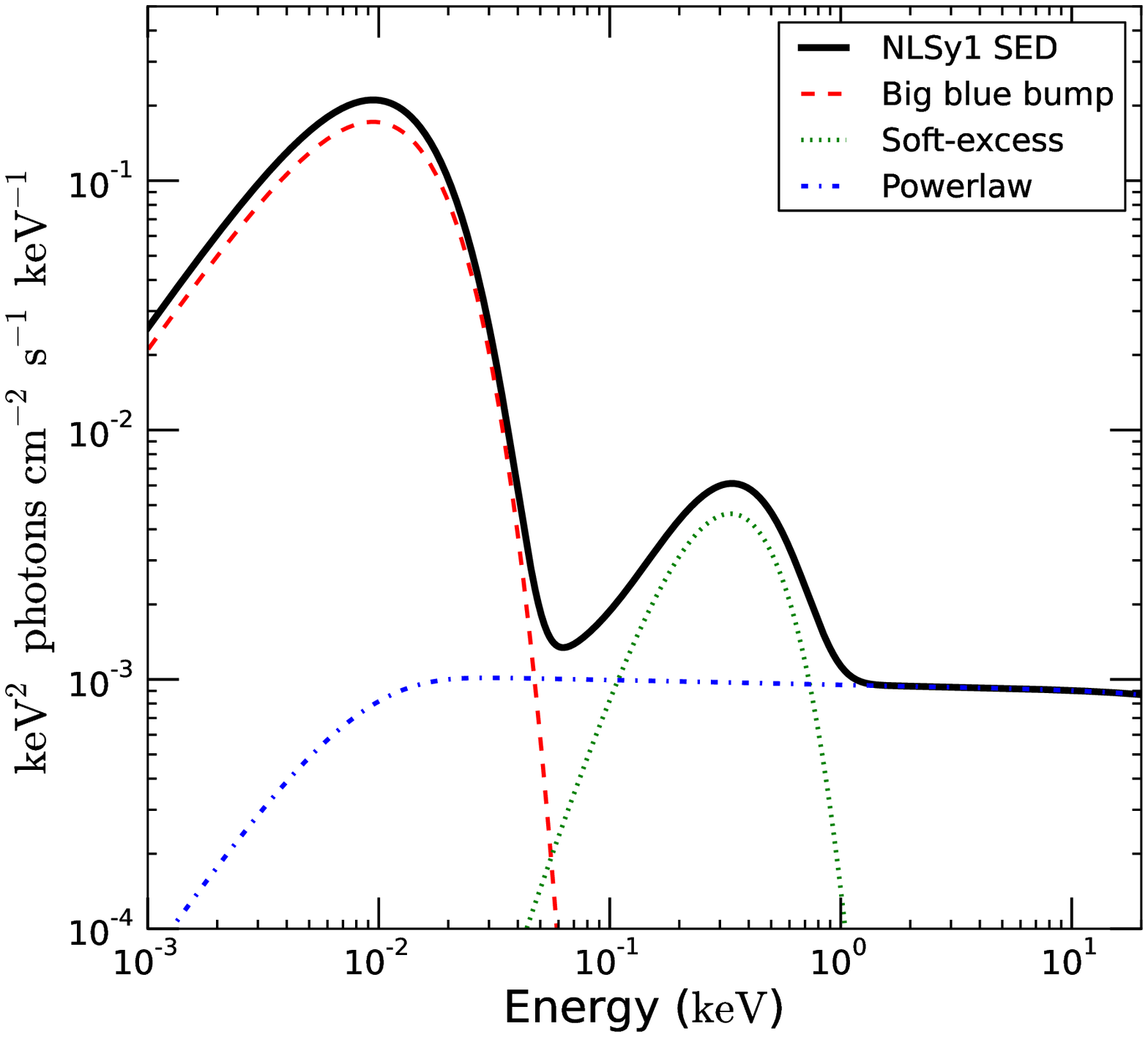}
  \includegraphics[width=8cm,angle=0]{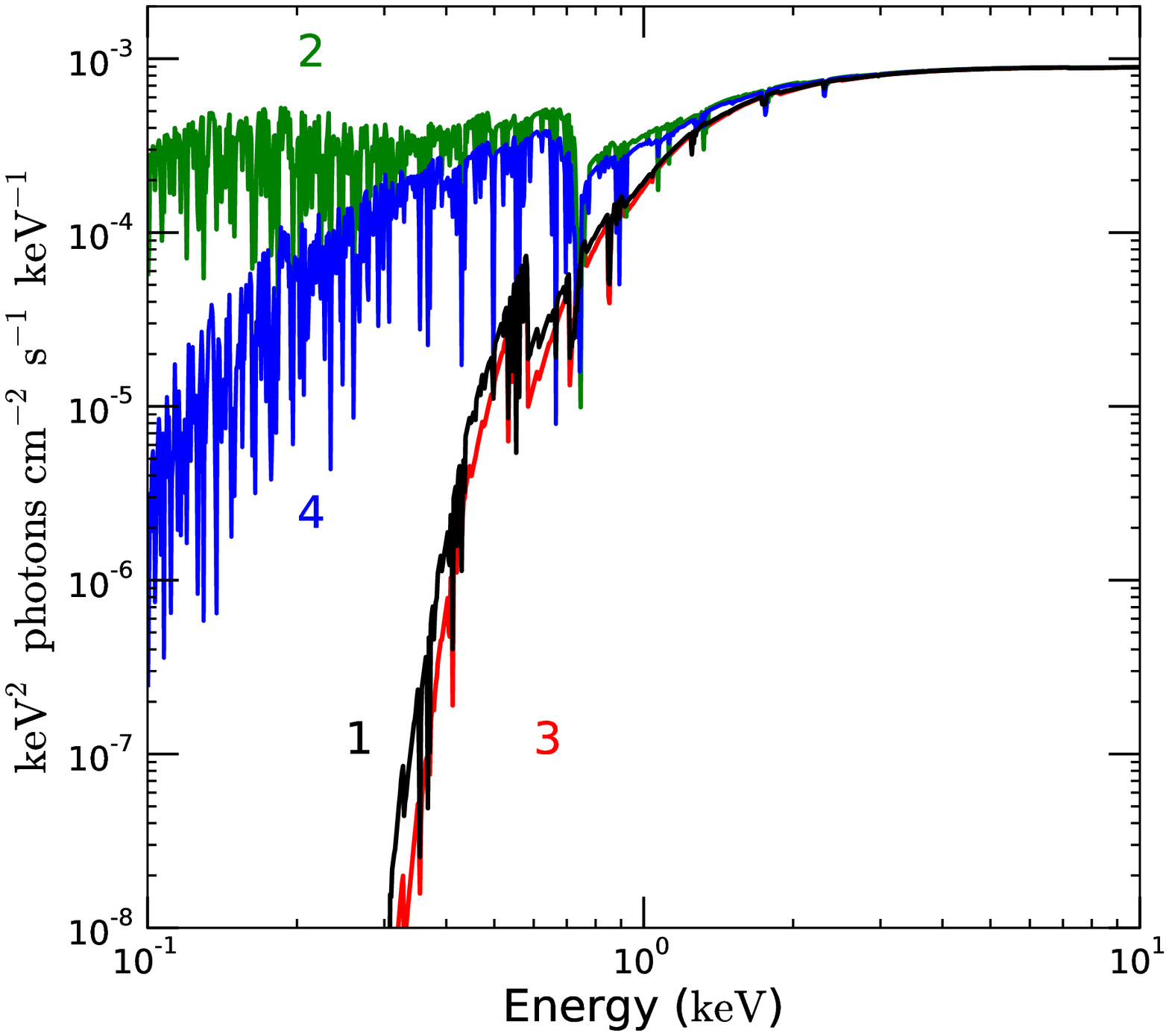}
  \caption{{\tt LEFT:} The black solid line is the NLSy1 ionising continuum as defined in section \ref{sec:sed}. The red dashed line, the green dotted line and the blue dashed-dotted line shows the three individual components of the continuum, the BBB, the SE and the powerlaw with a cut-off respectively. {\tt RIGHT:} The absorption features of a warm absorber
    cloud illuminated four ionising continua as described in section \ref{sec:sed} for the same ionisation parameter ($\xi=10 \xiunit$) and column density ($\nh=10^{22}{\rm~cm^{-2}}$) in the energy band of $0.3-10\kev$. The curves 1, 2, 3, and 4 denote the absorption from WA clouds illuminated by the four ionising continua, NLSy1, NLSY1 without BBB, NLSy1 without the SE, NLSy1 without the SE as well as the BBB.}
  \label{WA-sed}
\end{figure*}

\begin{figure}
  \centering \hbox {
    \includegraphics[width=5.6cm,angle=-90]{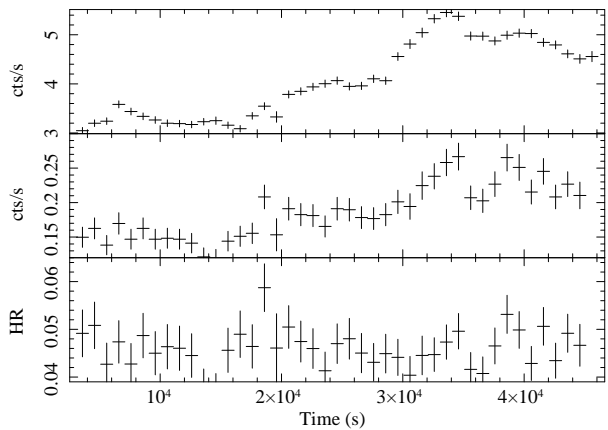}
  }
  \caption{The top and the middle panels are $0.2-2 \kev$ and $2-10 \kev$ background subtracted
    lightcurves respectively of the source IRAS~13349+2438, for the EPIC-pn data of the \xmm{} observation done in 2000. The bottom panel is the hardness ratio plot, which is seen to be constant during the observation. {Last 15 ks of the observation were not used for spectral analysis due to high particle background flaring}.}
  \label{HR}
\end{figure}

The above four ionising continua were used to create WA models using
the photoionisation code CLOUDY (version 08.00, see \cite{1998PASP..110..761F} for a description of CLOUDY), which uses an extensive atomic database to predict the
absorption and emission spectrum through and from a cloud. The clouds
are assumed to have a uniform spherical distribution around the
central source and are photoionised by the source. The geometry of the
cloud is spherical but we approximate it to a plane parallel slab by
making the distance of the cloud from the central source very large
compared to the thickness of the cloud.  CLOUDY performs the
simulations by dividing a cloud into thin concentric shells referred
to as zones. The thickness of the zones are chosen small enough for
the physical conditions across them to be nearly constant. For each
zone the simulations are carried out by simultaneously solving the
equations that account for ionisation and thermal balance. The model
predicts the absorption and emission from such clouds in thermal and
ionisation equillibrium. Following \cite{2006PASP..118..920P}, we
created WA models for each of the four continua described above.

To see how the warm absorption features of these four WA
models affect a spectrum, we have generated a powerlaw spectrum with
$\Gamma=2$ in the energy range $0.1-10\kev$ modified by the four WA models with the same column
$\nh=10^{22}{\rm~cm^{-2}}$ and the same ionization parameter $\xi=10
{\rm~erg~cm~s^{-1}}$. This was done using the {\it XSPEC} \citep{1996ASPC..101...17A} spectral fitting
package, where we have used a dummy response matrix to generate the
fake data. Figure \ref{WA-sed} left panel shows the full NLSy1 ionizing
continuum in solid black along with the three different constituents of the ionising continuum, the BBB, the SE and the powerlaw with a cut-off. The right panel of Fig. \ref{WA-sed} shows the absorption features of the warm absorber clouds corresponding to the four SEDs as described above. We see that ionizing continua with different shapes can produce clouds with different absorption features for the same ionization parameter.

This is physically understandable. For example, a continuum with a strong BBB will remove most of the  electrons from the atoms in the cloud capable of absorbing in the UV. On the other hand, a continuum with strong SE will remove all the electrons capable of absorbing in the soft X-rays from their shells. So a cloud illuminated by a strong SE is almost  `transparent' in the soft X-rays as it has no further electrons left to absorb the X-rays, as we find in  curve 2 in the right panel of Fig. \ref{WA-sed}. To
demonstrate these effects, we have performed a case study of the WA
properties of a bright NLSy1 IRAS~13349+2438 and discuss the effect of different parts of the ionising continuum on the WA properties.

\section{A case study of IRAS~13349+2438}
\label{sec:obs}

IRAS~13349+2438 is a nearby (z=0.107) bright radio
  quiet NLSy1 with a high bolometric luminosity ($\ge 10^{46}
  \lunit$). Previous X-ray studies have found the presence of strong
  WA, SE and a steep powerlaw spectrum with $\Gamma\sim 2.2$
  \citep{2003A&A...410..471L}. This is the source in which the Fe
  unresolved transition array (UTA) absorption features were detected
  by \cite{2001A&A...365L.168S} for the first time.  A multiwavelength study of this source by \cite{2013MNRAS.tmp..803L} has revealed the presence of multiple components of UV and X-ray warm absorbers. IRAS~13349+2438
  has been observed by \xmm on two ocassions in 2000 and once in
  2006. In the first observation in 2000, the
  X-ray spectrum showed the presence of a strong soft-excess and WA features
  \citep{2001A&A...365L.168S}. Figure \ref{HR} shows the hardness ratio along with the lightcurves in the soft and the hard band. We find that the  spectral shape has not varied during the observation. For these reasons we have chosen the source for our study.

\subsection{Observation and data reduction}
\label{subsec:obs}
We have used the archival \xmm data (id:0096010101) obtained from the
observation on 2000-06-20 for a total exposure of $45 \ks$. The
EPIC-pn and MOS cameras were operated in the small window mode using
the thin filter. The data were processed using SAS version 12 and the
latest calibration database, as available on ${3}^{rd}$ March 2012.

{The EPIC data were filtered using the standard filtering criterion as well as for particle background which
resulted in a net EPIC-pn exposure of $\sim 30 \ks$ in the EPIC-pn data which is similar to that obtained by \cite{2003A&A...410..471L}. We checked
the photon pile-up using the SAS task {\it epatplot} and found that
there was no noticeable pile-up in either EPIC-pn or MOS data. We
quote results based on EPIC-pn data due to its higher signal-to-noise
compared to the MOS data. We have used the good X-ray events (FLAG=0,
pattern=0). To extract the source spectrum we used a circular
region of $45{\rm~arcsec}$, centred on the centroid of the source. We
extracted the background spectrum from appropriate nearby circular
regions free of sources. We created the ancillary response file (ARF)
and the redistribution matrix file (RMF) using the SAS tasks {\it
  arfgen} and {\it rmfgen}. 
  We reprocessed the RGS data using the SAS task {\it rgsproc} and the OM data using the SAS task {\it omichain}. The OM
camera simultaneously observed IRAS~$13349$+$2438$ in the UVW2 filter along with EPIC and RGS
cameras. We have obtained the source flux at
$\rm 2120 \AA$ from the data and corrected it for Galactic extinction
following \cite{2011ApJ...737..103S} and
\cite{1998ApJ...500..525S}. We calculated the monochromatic flux at
$\rm 2120\AA$ from the AGN to be $\rm 3.4\times 10^{-15} \funit
\AA^{-1} $.

\subsection{X-ray spectral analysis}
\label{subsec:xray-analysis}
We begin with the spectral analysis of the broadband ($0.3-10\kev$)
EPIC-pn spectral data. The data were grouped with a minimum of 20
counts per energy bin and allowing 5 energy bins per
resolution element. This was done using the {\it specgroup} tool in
the SAS. We used ISIS version 1.6.2-12 \citep{2000ASPC..216..591H} for
our spectral fitting. An absorbed powerlaw fit to the spectrum in the $2-10 \kev$ band yielded a powerlaw slope
$\Gamma=1.96_{-0.09}^{+0.08}$ which is similar to that obtained by
\cite{2003A&A...410..471L}. There is possibly a very weak narrow Fe K$\alpha$
line and an Fe K absorption edge. We fitted the Fe K$\alpha$ emission
line with a narrow Gaussian and the fit improved by only $\dc=-4$ for
2 extra parameters from $\cd=127/125 \sim 1.02$, where dof stands for
degrees of freedom. The Fe K edge was fitted using an {\tt edge} model
in ISIS. The fit improved by $\dc=-28$ for 2 extra parameters. {The
best fit edge energy is $7.48_{-0.18}^{+0.12} \kev$, and the maximum
optical depth $\tau = 0.44\pm 0.22$, similar to that found by
\cite{2003A&A...410..471L}. The statistical improvement according to an {\tt F-test} upon addition of this component is $>99.9\%$. However the best fit $\cd=99/123$ suggests that the data may be over-modeled in this energy band.}

{ We extrapolated the model to the softer part of the spectrum and found
a prominent soft excess which was well described by a blackbody with
a best fit temperature of $\rm kT_{BB}=85\pm 2 \ev$ in the energy range $0.3-10\kev$, which is again
similar to that found by \cite{2003A&A...410..471L}. In Figure \ref{WA}, the left
panel shows clear
residuals of absorption features in the soft X-ray band, which are
mainly the signatures of absorption features. We used CLOUDY models to fit these absorption features.

\begin{figure*}
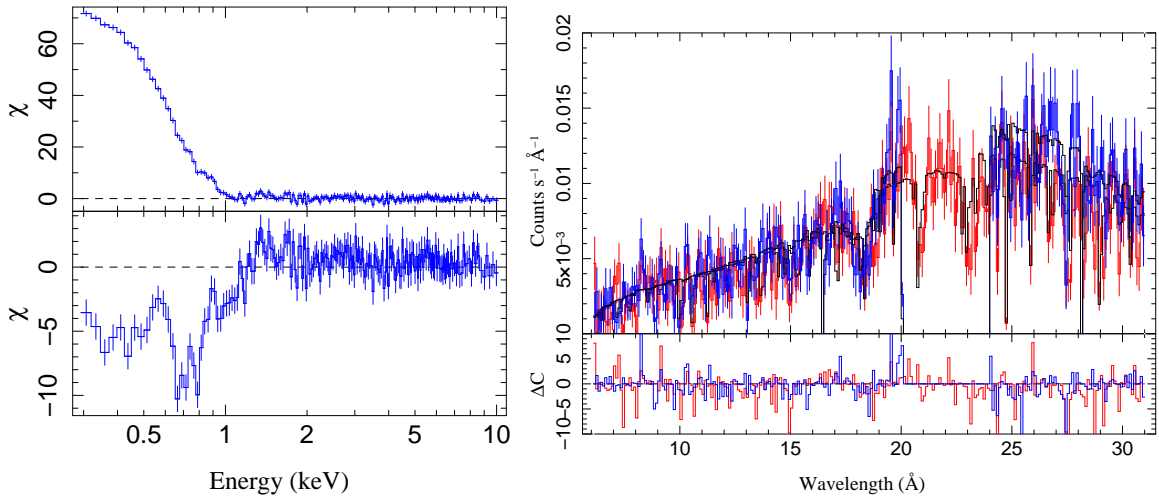

  \centering \hbox {
    \includegraphics[width=6.5cm,angle=-90]{multiplot.ps}
    \includegraphics[width=6.5cm,angle=-90]{IRAS13349-RGS-compare.ps}
  }
  \caption{{\tt LEFT :} EPIC-pn data residuals expressed as $\rm \chi={(data-model)}/{\sigma}$, where $\sigma$ denotes the statistical errors on the data. The
    upper panel shows the prominent soft-excess emission over an
    absorbed powerlaw fit in the $2-10 \kev$ range. The lower
    panel shows the presence of absorption features after we add a
    blackbody of temperature $\rm kT_{BB}=0.085\kev$ to the previous
    fit. {\tt RIGHT :} The RGS1 and RGS2 datasets in blue and red respectively, with the
    best fit model in black when fitted simultaneously with EPIC-pn data using
    the warm absorbers developed using the appropriate continuum (SE
    modeled with {\tt bbody}).}
  \label{WA}
\end{figure*}


 
\subsection{Constraining the broad band UV to X-ray continuum}
\label{subsec:xray-cont}

 The WA features in the X-rays consist of a number of absorption lines and edges of
varying strengths and are usually unresolved and
blended. In such cases, inferring X-ray continua from moderate
resolution X-ray spectra requires appropriate physical models of the
WA cloud. However, creating such a model requires the knowledge of the
ionizing continuum seen by the WA. This is somewhat of a circular
problem. We determine the continuum in the $0.3-10\kev$ band in the following way: We first use a generic AGN continuum to generate the CLOUDY WA table models. We use these WA models to fit the $0.3-10\kev$ EPIC-pn and the RGS data jointly to the X-ray continuum (Section \ref{sec:kirk}). The BBB is then derived following \cite{2013MNRAS.tmp..803L} (section \ref{subsec:BBB}). Finally we extrapolate the BBB and the X-ray continuum to the unobserved energy range of $13.6-300\ev$ and obtain the appropriate SED for the source. This SED is then used to create the WA table model and we obtain the best fit WA and continuum parameters from the joint fit of EPIC-pn and RGS data.

\subsection{The X-ray continuum}
\label{sec:kirk}
 In the first step we use the AGN continuum given by \cite{1997ApJ...487..555K} to generate the warm absorber table models in CLOUDY. The Kirk Korista continuum is given by} 
\begin{equation}
   f_{\nu}={\nu}^{\alpha _{uv}} \exp(-h\nu /kT_{\rm BBB}) \exp(-kT_{\rm IR}/h\nu)+\eta {\nu^{(1-\Gamma)}},
\end{equation} This consists of a power-law in the $1 \ev - 100 \kev$ band, and another steeper power-law in the UV whose upper exponential cutoff is parametrised with a temperature $T_{\rm BBB}$ and the lower infrared cut off by $T_{\rm IR}$. We used a typical Seyfert 1 X-ray powerlaw slope of $\rm \Gamma=2$, a UV bump blackbody temperature of $\rm T_{\rm BBB}=10^5\, K$ peaking at $\sim 10\ev$, and an $\alpha_{ox}=-1.2$. The UV spectral slope was assumed to be $\alpha_{uv}=-0.5$
\citep{1994ApJS...95....1E}. The CLOUDY table model was built using the methods described
in \cite{2006PASP..118..920P}. We varied
$\log ({\xi/\xiunit})$ from -2 to 4 and $\log(\nh/\cmsqi)$ from 19 to 24 and created a
multiplicative table model for the warm absorption. The cloud was
assumed to have solar metallicity. A hydrogen density of $ n_H \sim
10^9 \rm \cmcubei$ was assumed as the properties of the WA clouds are practically volume density independant in the range $\sim
10^2-10^{12} \rm \cmcubei$ \citep{1996ApJ...473..781N}. The table model was subsequently imported to the
ISIS package and was used to obtain the best fit $0.3-10\kev$
continuum parameters by carrying out a joint fit with the EPIC-pn and the RGS data. The best fit continuum parameters are: powerlaw slope $\Gamma=2$, and {\tt bbody} $kT_e=85\ev$. Two ionisation states of WA were detected with best fit parameters: for the lower
ionisation state $\log\xi=1.50$,
$\nhwa= 2\times 10^{21}\cmsqi$, and for the higher ionisation
state, $\log\xi=2.25$,
$\nhwa=2\times 10^{21}\cmsqi$. We report the best fit continuum and the WA parameters in Table \ref{kirk}.\\


\begin{figure}
  \centering \hbox {
    \includegraphics[width=8cm,angle=0]{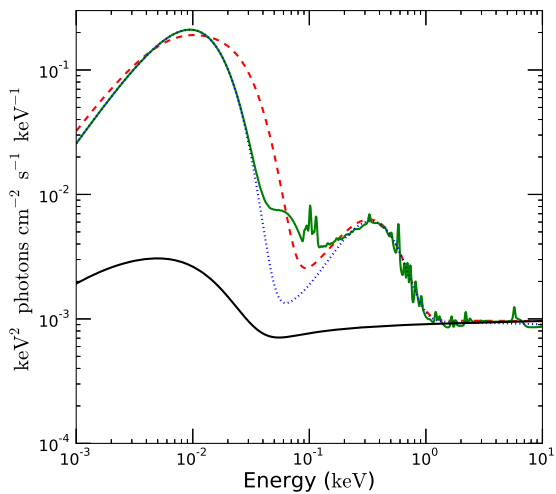}
  }
  \caption{{} The different {\it realistic} ionizing continua and their
    comparison with the generic {\it Kirk Korista} continuum shown by black solid curve. The
    dotted blue, dashed red, and solid green curves stand for SEDs where the soft excess is modeled by {\tt bbody}, {\tt optxagnf} and {\tt reflionx} models respectively. See Section \ref{subsubsec:DifftSe} for a discussion. 
  }
  \label{fig:real-model}
\end{figure}


\subsection{{Characterising the Big Blue Bump}}
\label{subsec:BBB}

The BBB in AGN is primarily
thought to arise from an optically thick but geometrically thin
accretion disk following \cite{1973A&A....24..337S}. We used the {\it
  diskbb} model \citep{1986ApJ...308..635M} in { ISIS} to define the
shape of the BBB, which requires the blackhole mass and distance to the
source for the normalisation to be determined. We have characterised the UV bump following \cite{2013MNRAS.tmp..803L}. They have derived the {\tt diskbb} parameters for the same source from the optical-UV data points obtained using Hubble Space Telescope (HST) after correcting for the intrinsic galactic reddening. The mass of the central massive
blackhole of IRAS~13349+2438 galaxy is estimated to be $\rm M_{BH}=10^{8.75}\msun$, an accretion efficiency of $20\%$ with respect to the Eddington rate and an inner radius of $\rm 10R_s$ (Schwarzschild radius). The normalisation of the {\tt diskbb} is obtained from these parameters and the luminosity distance of the source which is $\rm 483 \, Mpc$. Although the UV data used by \cite{2013MNRAS.tmp..803L} is non simultaneous to the \xmm{} observation we are working with, the fluxes by them are similar to those obtained by us using the OM UVW1 filter. 

\subsection{ The UV to Xray continuum and the Warm Absorber properties}
\label{subsec:appro}

We now develop a model ionising continuum as emitted by the central
active engine and as {\it seen} by the WA for the UV to X-ray energy band $1\ev-10\kev$. For the $0.3-10 \kev$ band
continuum, the best fit continuum model parameters obtained from the broad band EPIC-pn and
RGS data in Section \ref{subsec:xray-cont} is used, while the BBB is used as described in Section \ref{subsec:BBB}. In the unobserved energy range of $\sim 13.6-300 \ev$ the BBB meets the extrapolated $0.3-10 \kev$ continuum. 
Thus using the UV and X-ray
observations we have derived an appropriate continuum which is most likely
seen by the WA clouds in IRAS~13349+2438. {Fig. \ref{fig:real-model} shows this continuum as a blue dotted line}. 

 This ionizing continuum (hereafter `appropriate' continuum) is used to generate WA model
in CLOUDY which is then used to fit the EPIC-pn and RGS data
simultaneously. The best fit WA parameters obtained are: $\log\xi\sim1.75$,
$\nhwa\sim 2\times 10^{21}\cmsqi$ for the lower
ionisation state, and $\log\xi\sim3.25$,
$\nhwa\sim2\times 10^{21}\cmsqi$ for the higher ionisation
state (see Table \ref{kirk} for details). We
note that both the ionisation states are different from those obtained
using the WA model developed using the {\it Kirk Korista} ionizing continuum, although the continuum parameters are similar.


\subsection{Realistic ionising continuum and physical models for the Soft-excess}
\label{subsubsec:DifftSe}

In our attempt to create an appropriate ionizing continuum we have used a
blackbody to describe the soft-excess, which however is only a
phenomenological description and the resulting continuum may not be realistic. In the absence of observations in the $13.6-300\ev$, the best solution to derive the realistic continuum is to use physical models for the SE and the BBB. Though the nature of the SE is not clearly understood, it can be well described by two physical models (1) optically thick thermal Comptonisation ({\tt optxagnf}, in ISIS notation) and (2) blurred reflection from partially ionised accretion disk ({\tt reflionx}, in ISIS notation ).  

The {\tt optxagnf} model proposed by \cite{2012MNRAS.420.1848D} to model the soft-excess involves
Comptonisation of disk seed photons from a complex geometry. The AGN
spectral energy distributions in the UV to X-rays can be phenomenologically described by three main emission components: (1) the disk emission in the UV, (2) The soft X-ray excess emission from 
an optically thick, low temperature thermal Comptonizing plasma, and (3) and power law emission above $2 \kev$ from an optically thin, high temperature Comptonizing plasma. In the {\tt optxagnf} model these three main components
of the spectra are combined together assuming that they are all
ultimately powered by gravitational energy released in the accretion process. The {\tt optxagnf} model therefore simultaneously describes the UV as well as the X-ray spectra. However, we have only one data point in the UV from the OM UVW2 filter for the given observation which is not sufficient to constrain the model parameters. We created a set of fake datasets (in ISIS) in the UV ($1\ev-10\ev$) using the measured BBB described by \cite{2013MNRAS.tmp..803L} (see Section \ref{subsec:BBB}). We assumed a typical $5\%$ systematic errors on the UV data points. This UV data are simultaneously fitted with the EPIC-pn data to obtain the broadband UV to X-ray SED (see Figure \ref{fig:real-model}, red dashed line). The
model used is $ \rm \tt wabs\times WA1\times WA2\times edge \times
(optxagnf+Gaussian)$ in ISIS notation. The WA models used are those developed in the Section \ref{subsec:appro}. This continuum is now used to generate the WA model corresponding to the newly derived {\tt optxagnf} SED. We used this WA model to fit only the X-ray data ($0.3-10\kev$) with the model mentioned above and obtained the best fit WA and continuum parameters. The best fit parameter values of {\tt optxagnf}
model are $\rm \log(L/L_{Edd})\sim -0.76$, $\Gamma\sim1.99$ and the
temperature of the thermal optically thick Comptonizing electrons
responsible for SE are $ kT_e=101\ev$. The best fit WA parameters obtained are $\log\xi=1.95$,
$\nhwa= 2\times 10^{21}\cmsqi$ for the lower
ionisation state, and $\log\xi=3.52$,
$\nhwa=2\times 10^{21}\cmsqi$ for the higher ionisation
state (see Table \ref{real} and \ref{realistic} for details). We note here that the normalization of {\tt optxagnf} was frozen to one, since the flux is completely
calculated by the four parameters: the blackhole mass $\rm M_{BH}$,
the spin of the blackhole, the mass accretion rate $ L/L_{Edd}$, and
the luminosity distance of the source $ D_L$. 

The {\tt reflionx} model \citep{2005MNRAS.358..211R} describes the soft-excess as Compton reflection of hard X-ray photons from an ionised disk. It assumes a semi-infinite slab of optically thick cold gas of constant density, illuminated by a powerlaw producing a reflection component including the fluoresence lines from the ionised species in the gas. The XSPEC model {\tt kdblur} was convolved with the reflection model to account for the Doppler and gravitational effects. The $0.3-10\kev$ X-ray spectrum was initially fitted with a model {$\rm wabs \times WA1\times WA2 \times (powerlaw+kdblur(reflionx))$} in ISIS notation. The WA models used are those created using the appropriate continuum (Sec. \ref{subsec:appro}). The above model did not describe the data well ($\cd=305/177$). There were still some narrow discrete residuals in the soft X-rays. We added one more reflection component which is unblurred and is assumed to arise from a distant reflector. The fit statistic improved to $\cd=214/174$. We further tested the fit using more complex reflection models. We added one more blurred reflection component which arises from the same disk as the first one but lies on the outer side. We tied the outer radius of the first reflection component to the inner radius of the second component and let them free to vary. Since they arise from the same disk, we also assume that they have the same Fe abundance and same inclination angle with respect to the line of sight and we tie those two parameters as well. On addition of this component the fit improved to $\cd=201/171$ which is acceptable. No further reflection component was necessary. This X-ray continuum along with the BBB in the UV (described in Sec. \ref{subsec:BBB}) were used to create the broad band SED which was used to generate the WA table models corresponding to the {\tt reflionx} model. We used those WA models to fit the X-ray data. The EPIC-pn and the RGS data were simultaneously fitted with {$\rm wabs\times WA1\times WA2\times (powerlaw+kdblur(reflionx)+kdblur(2)(reflionx(2))+reflionx)$}. Table \ref{realistic} column 3 enumerates the best fit parameters obtained in the fit. Fig. \ref{fig:real-model} shows the three different ionizing continua derived when we used the different models to describe the soft-excess. We find that the WA parameters obtained for different soft excess models used, are similar within errors (see Table \ref{real}).  Hereafter these two continua will be referred to as the `realistic continua'.



\subsection{Effect of various continuum components on the warm absorber
  models}
\label{subsubsec:SwitchOff}

We constructed three new ionizing continua to investigate the individual
effects of the SE and the BBB by removing these components alternately from the appropriate ionizing continuum generated earlier (Sec. \ref{subsec:appro}). First we removed the diskbb component to get the NLSy1 continuum without the BBB, and created the WA model. Next we created another WA model using the NLSy1 continuum
without the SE. Finally we removed both the SE and the BBB from the NLSy1 continuum and created a WA model. The continua used are shown in Fig \ref{WA-sed}. We used these WA models based on different ionising continua to fit the absorption features in the EPIC-pn and RGS data. For each case we obtained a lower and higher ionisation state with different ionisation parameter. These are listed in Table \ref{switchoff}.


\begin{figure*}
  \centering
  {
    \includegraphics[width=8.6cm,angle=0]{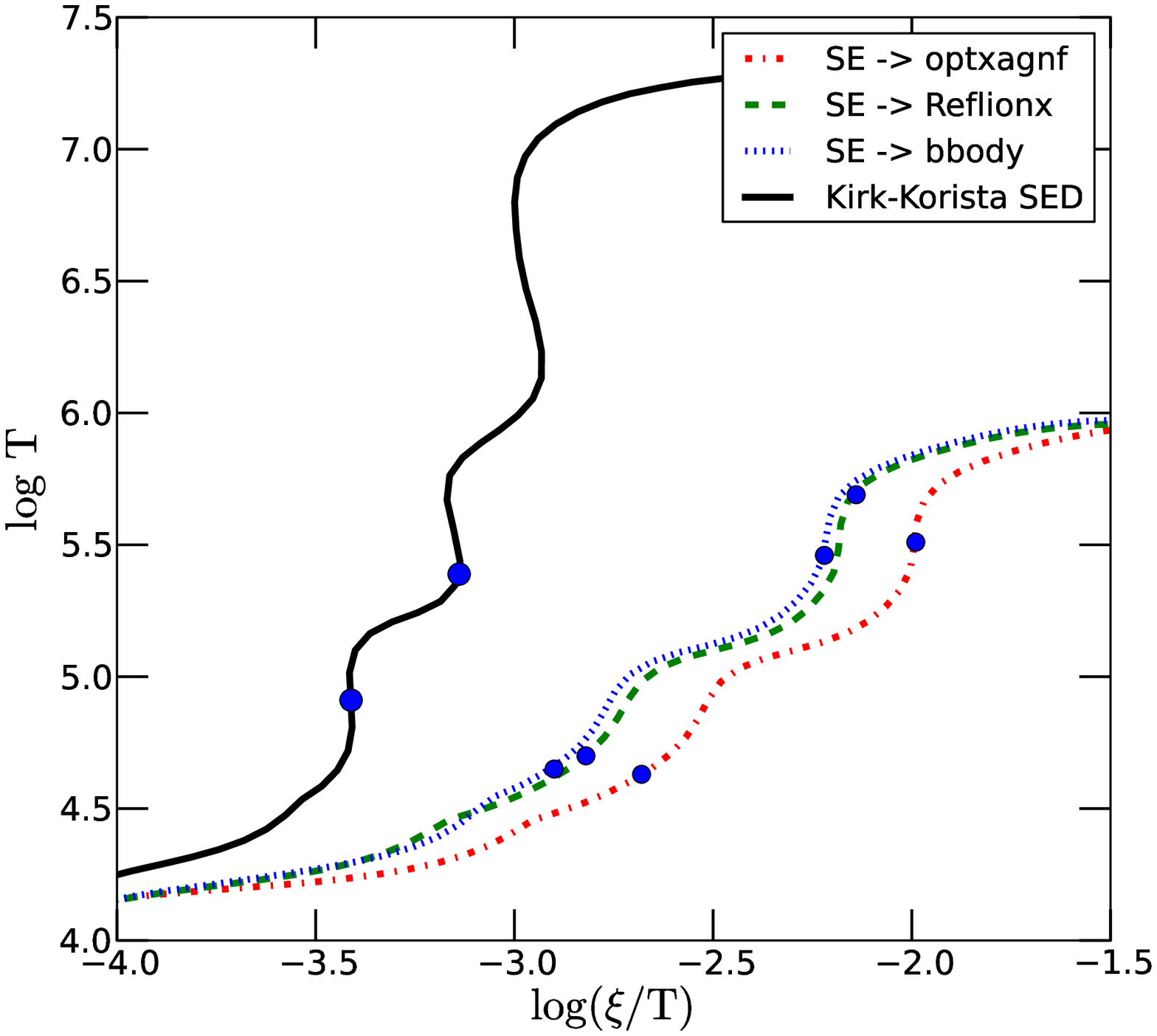}\includegraphics[width=8.6cm,angle=0]{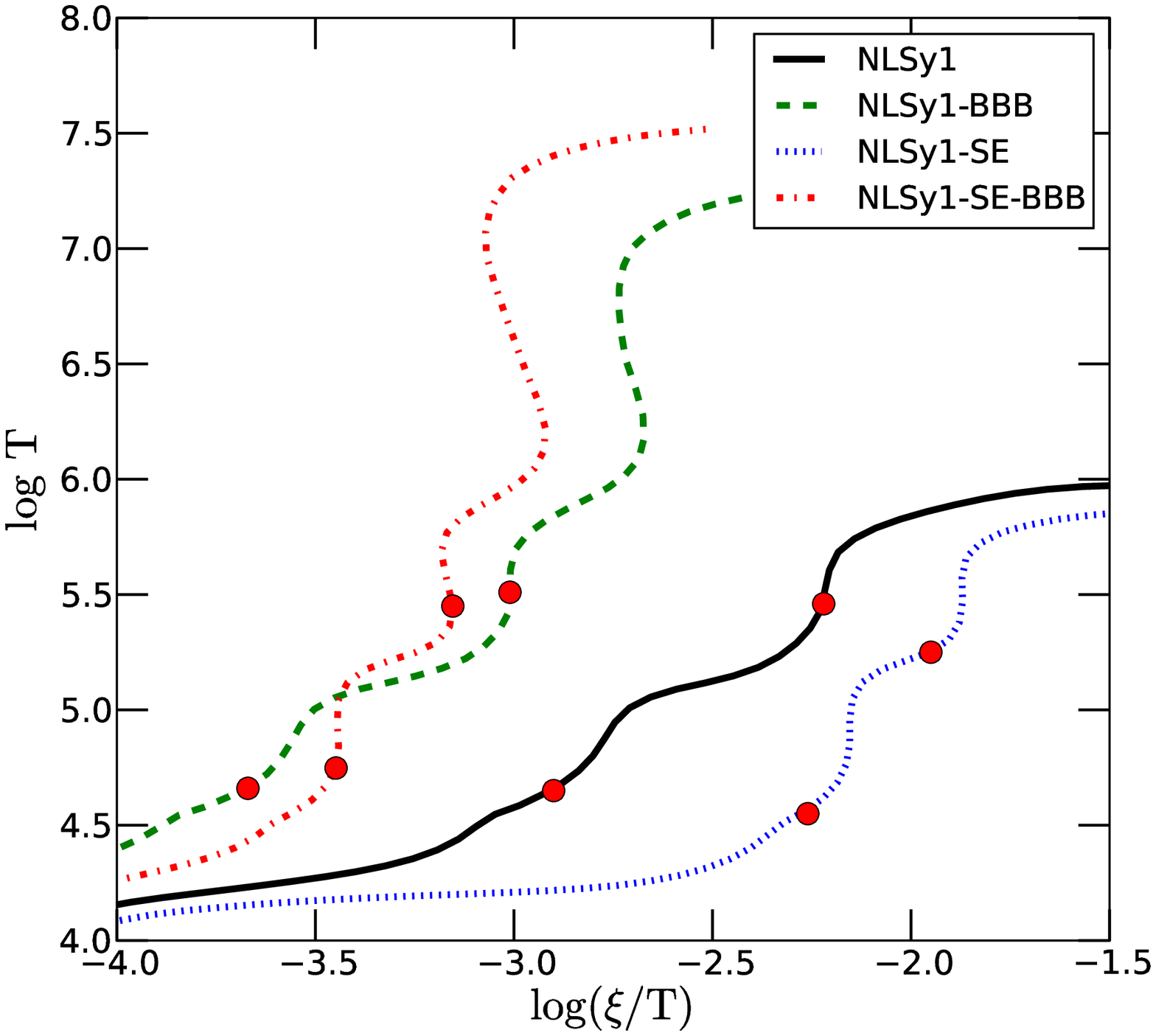}
  }
  \caption{{\bf LEFT:} The stability curves for the Kirk Korista
    continuum (solid black line) and for three other SEDs described in Sections \ref{subsec:appro} and \ref{subsubsec:DifftSe} where the SE is described by three different models ({\tt bbody}, {\tt optxagnf}, {\tt reflionx}). The filled circles
    are the two
    components of the best fit WA models for the respective
    continua. The WA components for the Kirk Korista continuum lie on
    the unstable part of the stability curve, but the WA components
    due to the realsitic continua lie on stable parts. {\bf RIGHT:} The stability curves
    for the test continua where certain components of an AGN ionizing
    continuum have been switched off, as discussed in Section
    \ref{subsubsec:SwitchOff} and Table \ref{switchoff}. The stability
    curve for the ionizing continuum where the BBB is switched off, SE is switched off, both SE and BBB are switched off
    are represented by the dashed green curve, dotted blue curve and dashed-dotted red curve. The
    corresponding best fit WA models are denoted by red points.}.
  \label{fig:Scurves}
\end{figure*}

\subsection{Stability curve analysis}

In the WA temperature range $10^4 - 10^{6.5}$ K, photoionization is
the main heating agent. Radiative recombination and line emission are
dominantly responsible for cooling the gas. The WA is assumed to be in
ionization and thermal equilibrium where the above physical
processes balance each other. 

The stability curve, which is a plot of the temperature T against the pressure $\xi/T$, is an effective tool that is often used to study the stability properties of warm absorbers. Every point on this curve represents a possible equilibrium state of the gas  \citep{1981ApJ...249..422K,1995ApJ...447..512K,1995MNRAS.273.1167R,2000A&A...354..411K,2001A&A...374..914K,2003ApJ...597..832K,2003ApJ...599..933N,2005ApJ...622..842K,2005ApJ...620..165K,2009MNRAS.393...83C,2012MNRAS.422..637C}. A point on a portion of the curve with positive slope corresponds to a state of stable thermal equilibrium because any small change in temperature will be
countered by the physical processes to leave its physical conditions
unchanged. However, if an absorber lies at a position on the curve with negative slope, then any small change in temperature will result in
runaway heating or cooling until the properties of the gas adjust to
reach stable configuration at higher or lower temperatures.

The stability curve for a given continuum is generated using CLOUDY by
stepping through different values of the ionisation parameter and
calculating the corresponding equilibrium temperature of the cloud. 

 The left panel of Figure \ref{fig:Scurves} shows
the stability curves for the Kirk Korista continuum and the three other continua developed for the source (SE modeled with {\tt bbody}, {\tt optxagnf} and {\tt reflionx}). The
Kirk-Korista continuum stability curve (solid black line) shows
distinct regions of stable phases (positive slopes) separated by
intermediate unstable phases (negative slopes). Such a stability curve
would predict discrete phases of WA which may be in pressure
equilibrium with each other. This is because $\rm \xi/T$ is essentially the ratio between radiation pressure and gas pressure if the distance from the ionizing source is the same for the different clouds of ionized gas. The solid blue points on the curve show the two components of the WA that were
derived from fitting the data. We find that both the components of
the WA fall in nearly unstable regions of the stability curve. On the other hand, the 
curve due to the two realistic continua ({\tt optxagnf} and {\tt reflionx}) predict mostly stable gas. Such curves correspond to a continuous distribution of temperature and pressure for the absorbing
gas. The WA components predicted by fitting the table models of these ionizing continua (see Table \ref{real}) are all thermally stable. Thus, for IRAS13349+2438,
the stability curve analysis points to the fact that the realistic ionising continua are the preferred ones for producing stable states of the warm absorbers. In the right panel of Figure \ref{fig:Scurves} we compare the stability
curves derived using the appropriate SED for IRAS~13349+2438 (see Sec. \ref{subsec:appro}) with stability curves obtained by dropping the BBB and the SE components respectively from the SED. We have shown in the figure with red circles the WA
components predicted by fitting the data with the WA models created using
the respective continua. We find that the BBB as well as the SE components are
crucial in establishing a largely stable curve and having the WA
components in the stable parts of the curve. 
\section{Discussion}

We have shown that ionizing continua with different shapes can result in different ionisation structures for a cloud even if they have the same value for the ionisation parameter $\xi$ \citep[see Fig. 1, see also][]{1999ApJ...517..108N,1995MNRAS.273.1167R}. The 13.6-300 eV range in the UV to X-ray SED of AGN is not accessible to us with our current state of the art instruments. This is the energy range whose photons are important for maintaining the ionisation balance of the WA clouds. Several authors in the past have used different techniques to overcome this difficulty. In many cases, the authors have used a single powerlaw ionising continuum with typical photon index of $\Gamma\sim 1.5-2$ in the UV to X-ray band \citep[e.g.,][]{ 1995MNRAS.273.1167R, 2011A&A...533A...1M,2011MNRAS.414.1965L,2012ApJ...745..107W}. In some cases a powerlaw connecting the two observed fluxes at $2\kev$ and $\rm 2500 \AA$ characterised by a slope $\alpha_{OX}$ has been used for the unobserved part of the continuum \citep[see e.g.,][]{1993ApJ...411..594N, 2008A&A...484..311L,2010A&A...515A..47N}. This method ignores the presence of the SE, which we have seen is important in maintaining the ionisation balance of the WA. Some authors have modeled the UV data with a powerlaw and joined the last UV data point, for instance, $\rm 2120\AA$ from the OM telescope, with the lowest available X-ray data point ($\sim 300 \ev$) to create the broad band SED \citep[see for e.g.,][]{2013ApJ...768..141G,2005ApJ...620..165K,2003ApJ...582..105Y}. \cite{2007ApJ...671.1284D} have treated the soft X-ray excess as disk blackbody emission and used a UV powerlaw to construct the ionising continuum.


 To constrain the spectral shape in the unobserved energy range, the best one can do is to use physical models to describe the BBB and the SE using the observed data and extrapolate them into the unobserved range. \cite{2013MNRAS.tmp..803L} have modeled the SE in the \chandra HETG spectrum of IRAS~13349+2438 with a physical model, {\tt nthcomp}, which involves Comptonisation of seed photons from an accretion disk by a thermal gas of electrons, and extrapolated the resulting continuum below $100\ev$ where it meets the BBB in the unobserved energy range. In the present case study of the NLSy1 galaxy IRAS~13349+2438 using \xmm{} data we have followed a similar procedure, but we have used two additional physical models to describe the SE - the intrinsic disk thermal Comptonisation {\tt (optxagnf)} and the blurred reflection from an accretion disk {\tt (reflionx)}. The {\tt optxagnf} model is similar to {\tt nthcomp} model for the SE but it also produces the BBB and the hard X-ray powerlaw self consistently. We find that these different models when extrapolated to the unobserved energy range predict different fluxes (see Fig. \ref{fig:real-model}). The model {\tt optxagnf} which simultaneously describes the X-ray and the UV data predicts greater flux in the $10-100\ev$ range compared to the {\tt bbody} model. The {\tt reflionx} model which required a separate {\tt diskbb} model to characterise the BBB predicts a greater flux in the energy range $20-100\ev$. The predicted fluxes in the $13.6-300\ev$ for the three cases {\tt optxagnf}, {\tt reflionx} and {\tt bbody} are $2.5\times 10^{-10}\funit$, $1.5\times 10^{-10}\funit$, and $1.4\times 10^{-10}\funit$, respectively. However the X-ray data we are working with cannot distinguish between the WA models created by the two physical SEDs. We find that the two different models lead to similar best fit WA parameters (see Table \ref{real}). The implication of these results is that the observed WA properties cannot be used to distinguish between the currently available physical models for the SE. 

The ambiguity in the derived WA ionisation parameters for a source due to the uncertainty in the SED means that we cannot determine the `unique' ionisation parameter for a given cloud for a given dataset. For example, when we use the appropriate SED (described in section \ref{subsec:appro}) for the source IRAS~13349+2438 we obtained the best fit ionisation parameters for the WA clouds $\log\xi\sim1.75$ and $\log\xi\sim3.25$, whereas the WA models developed using the SED when the BBB is switched off from the NLSy1 SED yielded completely different best fit WA ionisation parameters $\log\xi\sim0.99$ and $\log\xi\sim2.50$. \cite{2001A&A...365L.168S} studied the same observation of the source IRAS~13349+2438 and obtained two phase WA clouds. After modeling
the continuum, they applied the absorption components of
individual ions H and He-like C, N, O and Ne and Fe XVII-XXIV, where
each ion was treated as a separate component in the spectral fit. From
the observed distribution of the charge states of these ions, they
derived the average ionisation parameters using XSTAR: for the lower
ionisation state $\log\xi\le 1.0$, $\nhwa \sim 10^{21}\cmsqi$, and
outflow velocity is $\rm v \sim 400 \kms$, and for the higher
ionisation state $2.0 \le \log\xi \le 2.5$, $\nhwa \sim 10^{22}\cmsqi$
and outflow velocity is $\rm v \sim 0 \kms$. We found
  that both ionisation parameters are 
  consistent with those derived by us when we used the WA models based on the appropriate continuum without the BBB (see Table \ref{switchoff}). The ionisation parameters obtained by us using the WA model based on
  realistic continuum are higher than those obtained by
  \cite{2001A&A...365L.168S}, however they are similar to those obtained by \cite{2013MNRAS.tmp..803L}. \cite{2012A&A...542A..30M} have found different best fit WA parameters when they used different SEDs to generate the WA models. We found that the outflow velocity and the column density parameters for the WA models generated using different SEDs are statistically similar which is also found by \cite{2012A&A...542A..30M}.

The structure of the warm absorber clouds is still elusive. There is as 
yet no consensus on whether the clouds exist in clumpy discrete
phases or as continuos plasma. In two of the high
quality X-ray spectra of NGC~3783 \citep{2003ApJ...599..933N} and
NGC~5548 \citep{2003A&A...402..477S} the data required three
ionisation phases. It is not clear if these AGNs really host discrete
WA phases as the current data seem to suggest or the number of WA phases will
increase with increasing data quality, indicating a continuous
distribution of WA clouds. Various studies have however pointed towards a growing consensus on the discrete phases of WA \citep{2003A&A...403..481B,2008A&A...490..103S,2010A&A...518A..47R,2010A&A...514A.100M,2011A&A...534A..40E}. The stability curves provide a way to distinguish between the two scenarios. For the source IRAS~13349+2438 we have found that the effect of the BBB and the SE on the stability curves is to enable stable WA phases over a wide range of $\xi/T$ \citep[see also][]{1995MNRAS.273.1167R,1986MNRAS.218..457F}. { Therefore the continuum might influence the formation of specific phases.} \cite{1986MNRAS.218..457F} showed that the two phase discrete WA model for the source Mrk~841 is not valid as the stability curves exhibited more stable regions when they added a SE component to an existing simple powerlaw model in the X-ray continuum. We find that the stability curve generated using the Kirk Korista continuum which has no SE, has discrete stable and unstable phases. The best fit WA parameters for this continuum lie on the unstable portions of the curve. This may point towards a picture where the WA clouds exist in clumpy phases. On the contrary the stability curves generated using the realistic SED for IRAS13349+2438 with the BBB and SE (Fig. \ref{fig:Scurves}, left panel)  has only stable phases and the best fit WA parameters lie always on stable parts of the curves. This points to a continuos distribution of WA clouds as the available phase space in $\xi/T$ is largely stable. This can impact our understanding of the formation of the WA \citep{1995MNRAS.273.1167R}. \cite{2013MNRAS.tmp..803L} also point out that the presence of a strong UV and SE component creates more stable phases in the stability curve. \cite{2007ApJ...663..799H} had used only the X-ray continuum for the same source to generate stability curves and found unstable phases, which are not present when the SED by \cite{2013MNRAS.tmp..803L} is used. Therefore we find that the stability curves generated using the realistic continua prefer a continuous distribution of WA ionisation states. {However, as a caveat we may consider that the detection of two discrete WA components in the stable region of the S-curve may not always imply that there is a continuous distribution of ionization states in the absorber along the line of sight to this target. The S-curves built using the 'realistic' continua shows largely stable portions and allow for a continuos distribution of ionisation states and yet we detect only two discrete states. The fact that we see only 2 components with very different ionization degrees and nothing in between, may argue against the continuous-flow distribution of ionization states.}

\section{Conclusion}
In this paper we have investigated the effects of the shape of the
ionizing continuum on the warm absorber properties. The main results
are as follows.

\begin{itemize}

\item Ionizing continua with different shapes create different
  ionisation structures in WA clouds for the same ionisation parameter
  and column density of the cloud. The best-fit WA parameters obtained
  using WA models generated with different input ionizing continua for
  the Seyfert 1 galaxy IRAS~13349+2438 are different.

\item The determination of the accurate shape of the ionising continuum therefore becomes
  imperative for generating the WA models, which is however not
  possible due to the Galactic extinction in the range $13.6-300\ev$. The only way out is to characterise the BBB and the SE with physical models and extrapolate them in to the unobserved region of the SED.
  
 \item We developed realistic continua based on multiwavelength observations which consists of the SE and a powerlaw
  emission in the X-rays and the BBB in the UV. We found that the different physical models for SE (blurred Compton reflection and optically thick thermal Comptonisation) predict different fluxes in the unobserved energy range, but the current X-ray data quality does not allow us to distinguish between them using derived WA parameters. 

\item {The extent of stable regions in the stability curves is large for the realistic continuum which possibly indicates a continuous distribution of the WA clouds for this source.}

\end{itemize}

\begin{table*}[h!]
{\footnotesize
\centering
  \caption{The best fit model parameters for CLOUDY models using Kirk Korista and the realistic ionizing continua. \label{kirk}}
  \begin{tabular}{l l l l llll} \hline\hline 

Model  & parameters         &  Kirk Korista\tablenotemark{a,}\tablenotemark{c}  & Appropriate\tablenotemark{b,}\tablenotemark{c}   \\   
components    &                   &   WA model    & WA model   &  \\\hline \\

wabs   &  $\nh \times 10^{20}\, (\cmsqi)$     & $1.1(f) $&$1.1(f) $  \\ 
       &  (fixed)                                                                           \\ \\

Warm absorber   &$\log (\nhwa/\cmsqi) $  & $21.58_{-0.07}^{+0.10}$&$ 21.33_{-0.04}^{+0.09}$&  \\ \\
(CLOUDY)       &$\log(\xi/\xiunit) $     & $1.50_{-0.07}^{+0.08}$  & $1.75_{-0.09}^{+0.13}$    \\ \\
        &outflow-velocity \tablenotemark{d}      & $960_{-450}^{+480}$ & $960_{-450}^{+480}$  &         \\ \\
Warm absorber   &$\log (\nhwa/\cmsqi) $  & $21.31_{-0.16}^{+0.30}$&$ 21.36_{-0.25}^{+0.23}$& \\ \\
(CLOUDY)       &$\log(\xi/\xiunit) $     & $2.25_{-0.09}^{+0.13}$  & $3.24_{-0.16}^{+0.22}$ &        \\ \\
        &outflow-velocity \tablenotemark{d}      & $1170_{-600}^{+900}$&  $1200_{-600}^{+900}$&         \\ \\
bbody           &$\rm kT_{BB} \,(\ev)$          & $85_{-2}^{+1}$ & $87_{-2}^{+1}$       \\
                &norm                            &   $(12\pm 2)\tablenotemark{e} $ & $(12\pm 2)\tablenotemark{e} $   \\ \\
       
nthcomp &$\Gamma$    & $2.00 \pm 0.03$       & $ 2.00 \pm 0.02$&     \\ 
(powerlaw)  &norm& $(90\pm20)\tablenotemark{e} $   & $(90\pm20)\tablenotemark{e} $ &      \\ \\

Gaussian& norm                &  $ (0.16\pm0.13)\tablenotemark{c} $     & $(0.16\pm1.3)\tablenotemark{e} $   \\
        & Line E (rest)$\kev$ & $6.4\pm 0.007$       & $6.4\pm 0.005$                     \\
        & $\sigma(\ev)$       &   $0.001 \kev (f)$     &  $0.001 \kev (f)$                        \\ \\

edge    & Energy ($\kev$)& $7.48\pm 0.12$ &    $7.48\pm 0.12$                              \\
        & $\tau$      &  $0.44\pm 0.22$   &    $0.40\pm 0.22$                              \\  \\
$\rm C/dof$ &   &  $ 5940/5139$ & $5935/5139$  \\ \\ \hline \\
$\rm \cd$   &    & $213/181 $          &  $214/181 $          &         \\ 

(EPIC-pn fit only)&  &       &       &      \\ \hline \\

\end{tabular} \\ 
}
\tablenotetext{a}{WA table model generated using {\tt Kirk Korista} continuum.} 
\tablenotetext{b}{WA table model generated using the appropriate continuum with the SE modeled with a {\tt bbody}. See Sec. \ref{subsec:appro}}
\tablenotetext{c}{$(f)$ signifies frozen parameters.}
\tablenotetext{d}{The outflow velocity of the WA with respect to systemic velocity expressed in $\kms$.}
\tablenotetext{e}{These quantities are in the units of $10^{-5}$.}

\end{table*}


\begin{table}
{\footnotesize
\centering
  \caption{{The best fit warm absorber model parameters  with either the UV or the soft-excess part or both the parts of the realistic continuum switched off.} \label{switchoff}}
  \begin{tabular}{l l l l llll} \hline\hline 

WA    & Parameters & Model 1\tablenotemark{a}  & Model 2\tablenotemark{a}  & Model 3\tablenotemark{a} \\   
 Component & ---&   &   &  \\\hline \\

1. &  $\log (\nhwa/\cmsqi) $             &  $21.33^{+0.09}_{-0.05}$   &  $21.36^{+0.12}_{-0.12}$ &$21.38^{+0.12}_{-0.12}$  \\ \\

   &$\log(\xi/\xiunit) $&  $0.99_{-0.06}^{+0.08}$    & $2.29_{-0.09}^{+0.09}$   &$1.30_{-0.1}^{+0.1}$ \\ \\


2. &  $\log (\nhwa/\cmsqi)$              &  $21.49^{+0.11}_{-0.31}$   &  $21.44^{+0.12}_{-0.40}$ & $21.44^{+0.12}_{-0.40}$ \\ \\

   &$\log(\xi/\xiunit)$ &  $2.50_{-0.07}^{+0.12}$    & $3.30_{-0.09}^{+0.14}$   & $2.30_{-0.09}^{+0.14}$  \\ \\

\hline \\
$\rm C/dof$ &   &  $ 5935/5139$ & $5932/5139$ & $5932/5139$  \\ \\ \hline \\


\end{tabular} \\ 

\tablenotetext{a}{Model 1 stands for the BBB switched off from the NLSy1 SED as described in section \ref{sec:sed}. \\Model 2 stands for SE swiched off \\Model 3 stands for the BBB as well as the SE switched off.}

}
\end{table}

\begin{table}
{\footnotesize
\centering
  \caption{The best fit warm absorber model parameters when we use realistic ionizing continua to generate CLOUDY warm absorber table models. \label{real}}
  \begin{tabular}{l l l l llll} \hline\hline 

WA    &   Parameters &Model 1\tablenotemark{a}   & Model 2\tablenotemark{a}  & Model 3\tablenotemark{a}  \\   
component & --&  (bbody)  & (reflionx)  & (optxagnf)      \\\hline \\

1. &  $\log (\nhwa/\cmsqi) $ &  $21.33_{-0.04}^{+0.09}$  &  $21.40_{+0.04}^{-0.12}$  & $21.36_{+0.12}^{-0.09}$\\ \\

   &$\log(\xi/\xiunit) $&  $1.75_{-0.09}^{+0.13}$ & $1.88_{-0.12}^{+0.16}$ & $1.95_{-0.08}^{+0.09}$ \\  \\


2. & $\log (\nhwa/\cmsqi) $ &  $21.36_{-0.25}^{+0.23}$  &  $21.91_{-0.25}^{+0.13}$  & $21.59_{-0.15}^{+0.25}$\\ \\

   &$\log(\xi/\xiunit) $&  $3.25_{-0.16}^{+0.22}$ & $3.55_{-0.25}^{+0.11}$ & $3.52_{-0.15}^{+0.08}$ \\ \\

\hline \\
$\rm C/dof$ &   &  $ 5935/5139$ & $5927/5139$ &$5930/5139$ \\ \\ \hline \\

\end{tabular} \\ 
}

 \tablenotetext{a}{Models 1, 2 and 3 stands for cases when SE is described by {\tt bbody}, {\tt reflionx} and {\tt optxagnf}. See Sec. \ref{subsubsec:DifftSe} for details. }
\end{table}

\begin{table*}
{\footnotesize
\centering
  \caption{The best fit parameters for the simultaneous fit of the EPIC-pn and RGS data when the Soft-excess was modeled using physical models. \label{realistic}}
  \begin{tabular}{l l l l llll} \hline\hline 

Model  & paramters         &  Model 1\tablenotemark{a,}\tablenotemark{ b}    & Model 2\tablenotemark{a,}\tablenotemark{ b}    \\   
components   &                   &           &         &           \\\hline \\

wabs   &  $\nh \, (\cmsqi)\times 10^{20}$     & $1.1 (f) $&$1.1 (f)$ & \\ 
       &  (fixed)                                                                           \\ \\

WA1 and WA2    & Values quoted in Table \ref{real}& \\ \\

powerlaw&$\Gamma$       & $2.04 \pm 0.03$      & --- &          \\ 
        &norm           & $(5.9 \pm 2)\times 10^{-4}$ & --- &        \\ \\

Reflionx-1&Fe/solar       & $ 10_{-0.75}^{+0.0}$& --- &   \\ 
        & $\Gamma$      & $2.04 $  &  --- &\\ 
        &$\xi$          & $199_{-14}^{+3}$  & --- &      \\ 
	&norm          & $(1.43 \pm 0.19) \times 10^{-6} $& ---& \\ \\

Kdblur1  &index         &  $ 9.58_{-0.41}^{+0.50}$  & ---&   \\  
        &$\rm R_{in}(r_g)$   & $1.23_{-0.00}^{+0.07}$  & --- & \\
        &$\rm R_{out}(r_g)$   & $13.2_{-0.2}^{+0.5}$  & --- & \\
        &inclination(degrees)    &  $37_{-6}^{+3} $  & --- &\\ \\

Reflionx-2 &Fe/solar       & $10 \rm (tied)$& --- &    \\ 
        & $\Gamma$      & $2.04$                &  --- & \\ 
        &$\xi$          & $60_{-8}^{+21}$    & --- &      \\ 
	&norm          & $(4.0_{-1.0}^{+0.5} )\times 10^{-7} $& ---&  \\ \\

Kdblur2  &index         &  $3.55_{-1.5}^{+0.5}$ & &   \\  
        &$\rm R_{in}(r_g)$   & $13.2 \rm (tied)$                & --- & \\
        &$\rm R_{out}(r_g)$   & $400 (f)$                & --- & \\
        &inclination(degrees)    &  $37\rm (tied) $        &--- &\\ \\

Reflionx-3 &Fe/solar       & $0.75_{-0.18}^{+0.41}$& --- &    \\ 
        & $\Gamma$      & $2.04$                &  --- & \\ 
        &$\xi$          & $139_{-28}^{+76}$    & --- &      \\ 
	&norm          & $7.26_{2.20}^{+4.20} \times10^{-8} $& ---&  \\ \\

optxagnf & norm          &   ---&$ 1(f)$   &         \\
        & $\rm log(L/L_{EDD})$& ---&  $-0.755_{-0.004}^{+0.003}$  &                  \\
        & $\rm kT_e\, (\kev)$    &  --- &  $0.101^{+0.04}_{-0.02}$      &          \\
        & $\tau$        &  ---   &  $68_{-6}^{+5}$          &          \\
        & $\Gamma$      & ---   & $1.99\pm 0.02$  &                          \\
        & $\rm f_{pl}$  & ---   & $0.51\pm 0.02$ &   \\ 
        & $\rm r_{cor}\, (r_g)$   & ---   & $10\pm 2$       \\ \\

Gaussian& norm                &  ---  & $ 6.7_{+2.0}^{-2.2}\times 10^{-7}$   \\
        & Line E (rest)$\kev$ & --- & $6.4\pm 0.005$                     \\
        & $\sigma(\ev)$       &  ---  &  $0.001 \kev (f)$                        \\ \\
edge    & Energy ($\kev$)& ---&    $ 7.48\pm 0.20$                              \\
        & $\tau$      &  --- &    $0.46\pm 0.22$                              \\  \\

$\rm C/dof$ &   &  $ 5927/5139$ & $5930/5139$  \\ \\ \hline \\
$\rm \cd$   &    & $201/171 $          &  $223/176 $          &         \\ 

(EPIC-pn fit only)&  &       &       &      \\ \hline \\

\hline \hline
\end{tabular} \\ 
}
\tablenotetext{a}{Model 1=wabs*WA1*WA2*edge*(powerlaw+kdblur(Reflion)+kdblur2(Reflion2)),  Model 2 =wabs*WA1*WA2*edge*(optxagnf+Gaussian) }
\tablenotetext{b}{($f$) stands for frozen parameters} 

\end{table*}

$Acknowledgements:$  This work is based on observations obtained with XMM-Newton, an ESA science mission 
with instruments and contributions directly funded by 
ESA Member States and NASA. This research has made use of the NASA/IPAC Extragalactic Database (NED) which is operated by the Jet Propulsion Laboratory, California Institute of Technology, under contract with the National Aeronautics and Space Administration. SL is grateful to CSIR, Government of India
for supporting this work. {The authors are grateful to the anonymous referee for helpful comments and suggestions.}

\bibliographystyle{apj} \bibliography{mybib}

\end{document}